\documentclass[technote]{IEEEtran}
\usepackage[noadjust]{cite}
\usepackage{amssymb}
\usepackage{amsmath}
\usepackage{float}
\usepackage{subfloat}
\usepackage{subfig}
\usepackage{multicol}
\usepackage{lipsum}
\usepackage{cuted}
\usepackage[numbers,sort&compress]{natbib}

\usepackage{booktabs}

\usepackage{algorithm}			
\usepackage{algpseudocode}		
\usepackage{moreverb}

\usepackage{mathtools}
\makeatletter
\def\raremin{\@ifnextchar_\@rareminsub\@raremin}
\def\@rareminsub_#1{\min_{#1}\,}
\def\@raremin{\min\,}
\makeatother

\newcommand{\sign}{\text{sign}}

\ifCLASSINFOpdf
  \usepackage[pdftex]{graphicx}
  \DeclareGraphicsExtensions{.pdf,.jpeg,.png}
\else
  \usepackage[dvips]{graphicx}
\fi
\begin{document}
%

\title{Bulbar ALS Detection Based on Analysis of Voice Perturbation and Vibrato}

\author{\IEEEauthorblockN{Maxim Vashkevich\IEEEauthorrefmark{1}, \fbox{Alexander Petrovsky}\IEEEauthorrefmark{1}, \\Yuliya Rushkevich\IEEEauthorrefmark{2}}\\
\IEEEauthorblockA{\IEEEauthorrefmark{1}Belarusian State University of Informatics~and~Radioelectronics \\\
vashkevich@bsuir.by}\\
\IEEEauthorblockA{\IEEEauthorrefmark{2}Republican Research and Clinical Center of Neurology~and~Neurosurgery
}}


%


\maketitle

\begin{abstract}
On average the lack of biological markers causes a one year diagnostic delay to detect amyotrophic lateral sclerosis (ALS). To improve the diagnostic process an automatic voice assessment based on acoustic analysis can be used. The purpose of this work was to verify the sutability of the sustain vowel phonation test for automatic detection of patients with ALS. We proposed enhanced procedure for separation of voice signal into fundamental periods that requires for calculation of perturbation measurements (such as jitter and shimmer). Also we proposed method for quantitative assessment of pathological vibrato manifestations in sustain vowel phonation. The study's experiments show that using the proposed acoustic analysis methods, the classifier based on linear discriminant analysis attains 90.7\% accuracy with 86.7\% sensitivity and 92.2\% specificity.  

\end{abstract}


%
\IEEEpeerreviewmaketitle

\section{Introduction}

Amyotrophic lateral sclerosis (ALS) is neurodegenerative incurable disease, approximately 50\% of patients with ALS die within 30 months of symptom onset~\cite{Ingre-15}. There is no biological marker for ALS  and the diagnosis is made based on constellation of clinical observations. On average, it takes over a year to arrive at the diagnosis~\cite{Yunusova-13}. Using an acoustic analysis of voice and speech is a promising way to improve the process of ALS detection and monitoring of disease progression~\cite{Horwitz-16}. It becomes possible since bulbar motor changes (i.e., difficulty with speech or swallowing) are the first symptoms in approximately 30\% of persons with ALS~\cite{Spangler-17}. Almost all people with ALS display bulbar symptoms at later stages. It is important that abnormal acoustic parameters of voice were also demonstrated in ALS subjects with perceptually normal vocal quality on sustained phonation~\cite{Tomik-10}.

The need for development voice and speech diagnostic tools has motivated recent work on elaboration of special acoustic analysis methods. One important aspect of current research related to neurological disease detection (such as Parkinson's and ALS) is that they targeted to use smartphones and tablets to record the voice using standard microphone in the different home environments~\cite{Benba-16,Norel-18}. Advancement in this direction can lead to creation of easy to use self- and tele-monitoring and disease detection means.

In previous studies different approaches and speaking tasks are used for ALS detection. One general approach that tends to solve the problem of differential diagnosis consists in classification of dysarthria type~\cite{Guerra-03,Mujumdar-10}. The main difficulty of this approach is that it requires the collection of representative dataset for all types of dysarthria. More simple approach to ALS detection is based on using dataset containing pathological and normal voice/speech samples~\cite{Spangler-17,Norel-18,An-18,Vashkevich-18}. In several studies detection of ALS is performed using kinematic sensors~\cite{Bandini-17,Spangler-17} to model articulation and measure prosodic elements such as vowel duration or speaking rate~\cite{Yunusova-13}. {\it Running speech} test was used in~\cite{An-18,Norel-18,Vashkevich-18} as a basis for ALS detection. In~\cite{An-18} representational learning approach based on convolutional neural networks (CNN) was applied for ALS detection. As a low-level features filterbank energies with its 1st and 2nd derivatives were used. The best sample-level accuracy was obtain by time-domain CNN (71.6\% sensitivity and 80.9\% specificity). In~\cite{Norel-18} huge feature set obtained with openSMILE toolkit was analysed, the mel-frequency cepstral coefficients (MFCCs) and features based on RASTA filtering were found the most informative. However, the support vector machine (SVM) classifier based on liner kernel presented in~\cite{Norel-18} has moderate accuracy (79\% for males and 83\% for females). In~\cite{Vashkevich-18} joint analysis of the vowels /a/ and /i/ extracted from running speech test was used for ALS detection. Classifier based on linear discriminant analysis (LDA)  reported in~\cite{Vashkevich-18} has 88\% accuracy (90.5\% sensitivity and 84.6\% specificity).  A {\it diadochokinetic task} (rapid repetitions of syllables) was used for automatic detection of ALS in~\cite{Spangler-17}. Fractal jitter introduced in~\cite{Spangler-17} along with MFCCs and articulatory features allows to obtain classifier based on extreme gradient boosting (XGBoost) with 90.2\% accuracy (94.5\% sensitivity and 85.1\% specificity). Our work extends prior efforts in following aspects, it 1) verifies the suitability of the most simple {\it sustain vowel phonation} test (to the best knowledge of the authors there is no works dedicated to ALS detection based on sustained phonation test), 2)~adapts acoustic analysis methods to work with non-studio voice recordings obtained using smartphone, 3)~proposes new acoustic feature called pathological vibrato index (PVI) that considerably increases the detection accuracy. 

\section{Voice perturbation analysis}

The perturbation measures jitter and shimmer and their variants are often use to assess the voice function~\cite{Baken-2000}. All of them require preliminary segmentation of the voice signal into fundamental periods. This problem is very similar to $f_o$ extraction but has its own specifics.

\subsection{$F_o$ extraction based on waveform matching}
In perturbation analysis it is conventional to use event detection methods (peak-picking or waveform matching) for $f_o$ extraction~\cite{Titze-93,Baken-2000}. The main reason of this is that objective of perturbation analysis is cycle-to-cycle variation of vocal fold vibration and therefore short-term average methods (when several cycles are analysed) are not applicable. In waveform matching (WM) method entire waveshapes are matching across adjacent cycles. The difference between periods is decided by a least square difference between adjacent cycles~\cite{Titze-93}. Currently waveform matching is concerned as the method that allows to obtain good results for voice perturbation analysis~\cite{Boersma-09}. In the widely used PRAAT voice analysis system there is an implementation of the WM method~\cite{Boersma-02,Boersma-09}.

\begin{figure}[t]
  \centering
  \includegraphics[width=\linewidth]{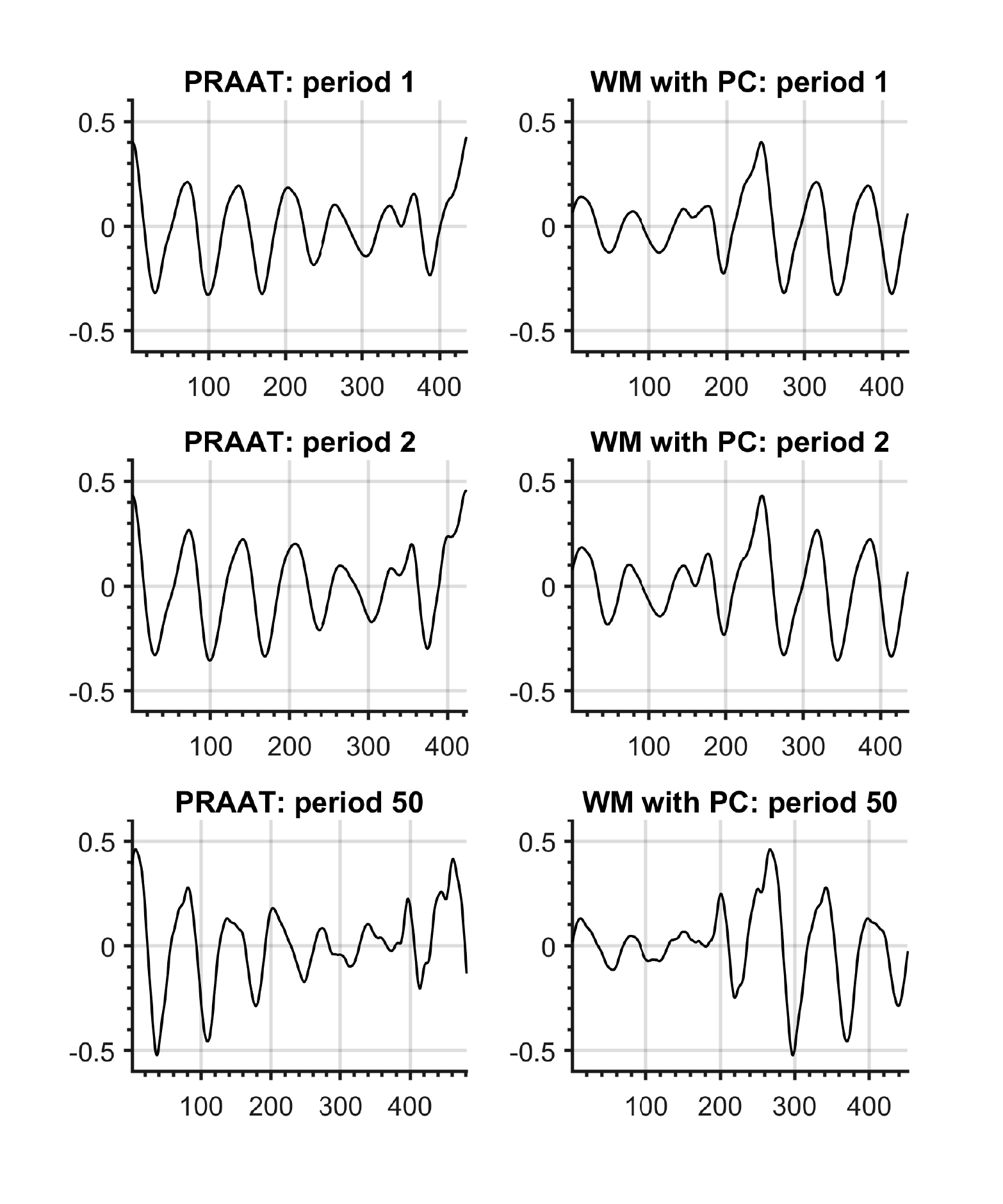}
  \caption{Phonatory cycles of pathological voice extracted using PRAAT and WM with phase constrain}
  \label{fig:praat_periods}
\end{figure}

Waveform matching method exploit the fact that current period is closely related to previous period. The drawback of this approach is that if an error appears in detecting current period it will affect all following periods. In practice it results in phase shift if we compare cycles distant one from another. The example of this phenomenon can be seen in Fig.~\ref{fig:praat_periods}. Period \#1 is almost the same as period \#2, however if we compare it with period \#50 the phase shift can be seen. The risk of errors in the WM method increases in the case of using non-studio recordings obtained using smartphone, that can contain environmental noise fluctuation. In this paper we propose a method aimed at eliminating these drawbacks.

\subsection{Waveform matching with phase constrain}
The proposed procedure starts by pitch estimation $f_o(n)$ using any general purpose method. The estimated $f_o$ contour is used as a ``reference'' signal for further period detection. In this work we used IRAPT estimator~\cite{Azarov-12} on this stage. The first period is detected using cumulative sum
\begin{equation}
  \Phi(n) = \sum_{k=1}^{n}\omega_o(k),
  \label{eq:Phi_n}
\end{equation}
where 
\begin{equation*}
  \omega_o(k)=2\pi f_o(k) / F_s,
\end{equation*}
where $f_o(k)$ -- fundamental frequency contour in Hz, $F_s$ -- sampling frequency. Starting with $n=1$ and incrementing $n$ function (\ref{eq:Phi_n}) is calculated until $\Phi(n)$ exceeds $2\pi$, after that first period is set equal to $n-1$.

The following periods are detected using two-stage procedure: 1) detecting approximate period boundary using phase function (\ref{eq:Phi_n}) and 2) refining of boundary such that mean absolute error between two adjacent waveforms is minimal. The \texttt{Matlab}-implementation of the proposed method is given below.

\begin{verbatim}
function T0 = WM_phase_const(s, w0)
% Initialization
Ln = length(s);
Phi_n = 0;	% phase from eq.(1)
n = 0;      % sample index
Nc = 1;     % current cycle number

% Searching first period
while (Phi_n<2*pi)
  n = n + 1; 
  Phi_n = Phi_n + w0(n);   
end
      
T0(Nc) = n - 1;
T0_prev = s(1:T0(Nc));
cur_T0 = 1;	% length of current period
Phi_n = Phi_n - 2*pi;

% Search cycle
while (n<(Ln-T0(Nc)))             
 if (Phi_n>2*pi)                                        
  cur_T0 = cur_T0 - 1;
  % find better offset
  offset = round(cur_T0*0.1);
  b = -offset:offset;	% bias
  for k=1:length(b)
    T0_cur = s(n-1-T0(Nc)+1+b(k):n-1+b(k));            
    err(k) = mean(abs(T0_prev - T0_cur));
  end
  [~,min_ind] = min(err);
  n = n - 1 + b(min_ind);       
  Nc = Nc + 1;
  T0(Nc) = cur_T0 + b(min_ind);
  T0_prev = s(n-T0(Nc)+1:n);
                 
  cur_T0 = 0;	% Reset running variables
  Phi_n = 0;
  else
    n = n + 1;
    Phi_n = Phi_n + w0(n);
    cur_T0 = cur_T0 + 1;
  end    
 end
end
\end{verbatim}

An example of application of WM method with phase constrain to pathological voice analysis is given in Fig.~\ref{fig:praat_periods}. It can be seen that cycles extracted using proposed method are more phase synchronised compared to conventional WM.

\subsection{Amplitude and frequency perturbation}
\subsubsection{Jitter}
\begin{figure}[t]
  \centering
  \includegraphics[width=0.7\linewidth]{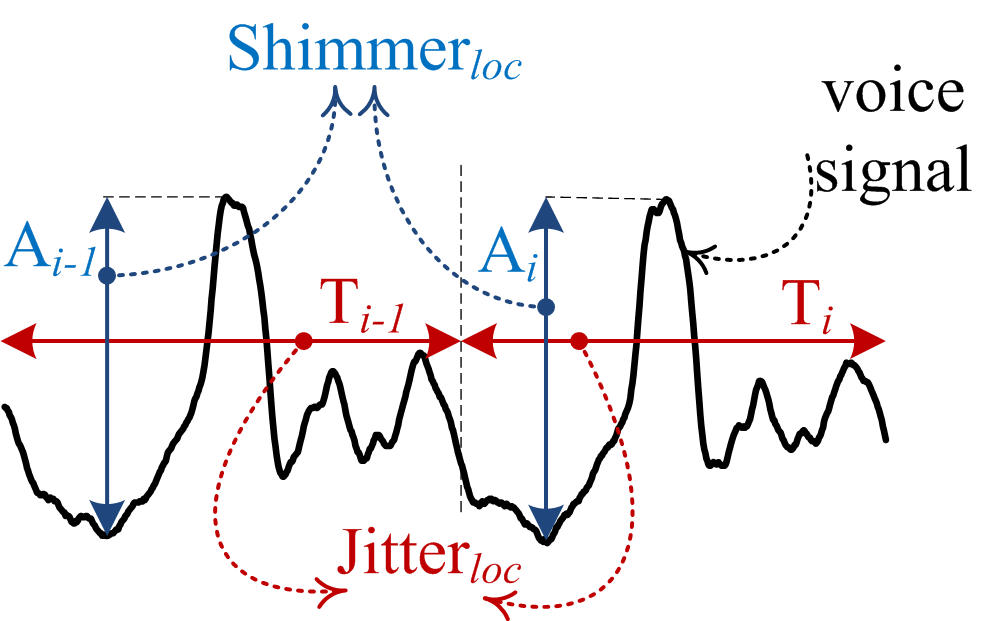}
  \caption{Jitter and Shimmer concepts illustration}
  \label{fig:jitter_shimmer}
\end{figure}

Jitter (frequency or period perturbation) is the variability of the fundamental period from one cycle to the next (see Fig.~\ref{fig:jitter_shimmer}). As far as jitter estimates short-term variation it can not be accounted to voluntary changes in $f_o$. Therefore jitter is intended to provide an index of the stability of the phonatory system. 
High level of jitter results from diminished neuromotor and aerodynamic control~\cite{Baken-2000}. 
The jitter has been used as an indicator of the voice quality that characterizes the severity of dysphonia~\cite{Castro-13}. 
Mathematically jitter ($J_{loc}$) is defined as average difference between consecutive periods, divided by the average period:
\begin{equation}
  J_{loc} = \frac{\textstyle \frac{1}{N-1}\sum\limits_{i=1}^{N-1}|T_0(i)-T_0(i+1)|}{
  \textstyle \frac{1}{N}\sum\limits_{i=1}^{N}T_0(i)}\times 100,
  \label{eq:Jit_loc}
\end{equation}
where $N$ is the number of extracted periods. 

Jitter can be also estimated as relative average perturbation (RAP) to quantify whether the period duration is smooth over three adjacent cycles~\cite{Baken-2000}:
\begin{equation}
  J_{rap} = \frac{\displaystyle \frac{1}{N-1}\sum_{i=2}^{N-2}\Bigl|T_0(i)-\frac{1}{3}\sum_{k=i-1}^{i+1}T_0(k)\Bigr|}{
  \displaystyle \frac{1}{N}\sum_{i=1}^{N}T_0(i)}\times 100.
  \label{eq:Jit_rap}
\end{equation}
In this study period perturbation quotient (PPQ) is also used to quantify the variability of the of the pitch period evaluated in five consecutive cycles:
\begin{equation}
  J_{ppq5} = \frac{\displaystyle  \frac{1}{N-4}\sum_{i=3}^{N-2}\Bigl|T_0(i)-\frac{1}{5}\sum_{k=i-2}^{i+2}T_0(k)\Bigr|}{
  \displaystyle \frac{1}{N}\sum_{i=1}^{N}T_0(i)}\times 100.
  \label{eq:Jit_ppq}
\end{equation}

\subsubsection{Shimmer}
Shimmer is a measure that characterizes the extent of variation of expiratory flow during the phonation. Basic shimmer measure ($S_{loc}$) is defined as average absolute difference between the amplitude of consecutive periods, divided by the average amplitude~\cite{Baken-2000}:
\begin{equation}
  S_{loc} = \frac{ \frac{1}{N-1}\sum\limits_{i=1}^{N-1}|A(i)-A(i+1)|}{
   \frac{1}{N}\sum\limits_{i=1}^{N}A(i)}\times 100,
  \label{eq:Sh_loc}
\end{equation}
where $A(i)$ is the amplitude of the $i$-th period.
$S_{loc}$ fall under influence of long-term changes in vocal intensity. To eliminate the effects of amplitude ``drift'' in order to get a truer index of underlying shimmer it has been suggested to measure amplitude perturbation quotient (APQ):

\begin{equation}
  S_{apqL} = \frac{\textstyle \frac{1}{N-L+1}\sum\limits_{\scriptscriptstyle i=1+\frac{L-1}{2}}^{\scriptscriptstyle N-\frac{L-1}{2}}\Bigl|A(i)-\frac{1}{L}\sum\limits_{\scriptscriptstyle k=i-\frac{L-1}{2}}^{ \scriptscriptstyle i+\frac{L-1}{2}}A(k)\Bigr|}{
  \textstyle \frac{1}{N}\sum_{i=1}^{N}A(i)}\times 100.
  \label{eq:Sh_apq}
\end{equation}
where $A(i)$ is the amplitude of the $i$-th period. The most commonly used window size $L$ are 3, 5 and 11~\cite{Orozco-16}.

\subsection{Vibrato analysis}
Vibrato is a rapid, and regular fluctuation of the $f_o$ that arises during sustained vowel phonation. A direct comparison ensures that there is a significant difference between the vibrato frequency in patients with ALS and healthy controls. 

The estimation of the extent of pathological changes in vibrato is based on the observation that for healthy voices vibrato lies in range of 5-8 Hz~\cite{Nakano-06} while for ALS patients characterized by presence of high-frequency components in 9-14~Hz range~\cite{Aronson-92}. In this study we use the following method of estimating \textit{pathological vibrato index} (PVI):
\begin{enumerate}
\item Estimation of $f_o(m)$ contour with 5~ms time step using IRAPT algorithm~\cite{Azarov-12};
\item Normalization of $f_o$ contour: $$f_o'(m)=\frac{f_o(m)}{\mathrm{mean}(f_o)};$$
\item Bandpass filtering of $f_o'(m)$ using 3-th order Butterworth IIR with pass band $[9,\;14]$ Hz; 
\item Amplitude spectrum $A_{f_o}(f)$ estimation using Welch's method with windows of 1 sec length and 95\% overlap;
\item Calculation of pathological vibrato index:
\begin{equation}
  \mathrm{PVI} = \sum_{f\in [9,\; 14] \textit{ Hz}}A_{f_o}(f).
  \label{eq:PVI}
\end{equation}
\end{enumerate}

\begin{figure}[t]
  \centering
  \includegraphics[width=\linewidth]{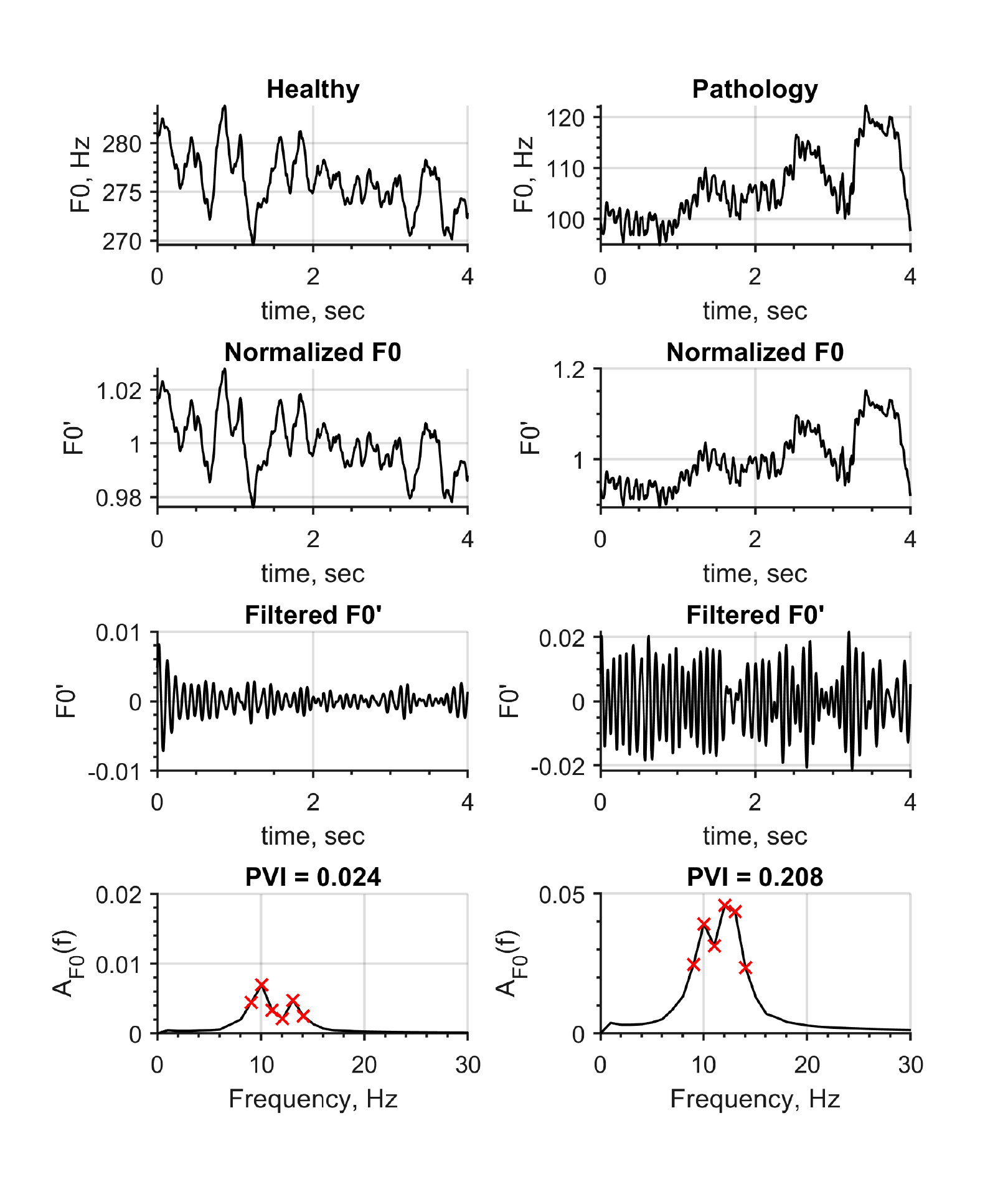}
  \caption{Left column: normal case, right column: pathological case. Rows from the top: Extracted $f_o$, normalized $f_o$ contour, IIR filtered $f_o$ contour, amplitude spectrum $A_{f_o}(f)$, amplitudes used for PVI calculation are indicated by red x-marks}
  \label{fig:pvi_steps}
\end{figure}

Fig.~\ref{fig:pvi_steps} shows the steps of the PVI calculation for a healthy speaker and ALS patient with moderate bulbar dysfunction.

\section{Classification}
For discriminating between the two classes of normal and pathological cases two widely spread machine-learning algorithms: linear discriminant analysis (LDA) and k-Nearest Neighbors (k-NN) were used.

\subsection{Linear discriminant analysis}
The LDA is an approach which uses linear hyperplane (decision surface) to separate the dataset into two classes~\cite{Cristianini-04}. In LDA the following classification function is used
\begin{equation*}
f(\mathbf{x})=\sign\bigl( \bigl<\mathbf{w},\mathbf{x} \bigr> + b\bigr),
\end{equation*}
where $\mathbf{x}$ -- feature vector, $b$ -- bias and $\mathbf{w}$ -- weight vector that defines decision surface and is chosen according to Fisher criterion to maximize the quotient
\begin{equation*}
J(\mathbf{w})=\frac{(\mu_{\mathbf{w}}^{+}-\mu_\mathbf{w}^{-})^2}{\bigl(\sigma_\mathbf{w}^{+}\bigr)^2+\bigl(\sigma_\mathbf{w}^{-}\bigr)^2},
\end{equation*}
where $\mu_{\mathbf{w}}^{+}$ and $\mu_{\mathbf{w}}^{-}$ are the mean of the projections of the positive and negative samples onto direction $\mathbf{w}$, $\sigma_{\mathbf{w}}^{+}$ and $\sigma_{\mathbf{w}}^{-}$ are the corresponding standard deviations.

\subsection{k-NN classifier}
The idea of k-NN classification consists in assigning to a new data sample the label by taking into account $K$ nearest samples in the training set. In this work the following approach was used: for new data sample $K$ nearest points $\mathbf{x}_{1\dots K}^{+}$ form positive class and $K$ nearest points $\mathbf{x}_{1\dots K}^{-}$ from negative class of training set were determined. Then simple voting technique based on {\it distances weighting}~\cite{Flach-2012} is applied to assign the label to the input sample $\mathbf{x}$:
\begin{equation*}
f(\mathbf{x})=\sign\Bigl( \sum_{k=1}^{K}\frac1{d(\mathbf{x},\mathbf{x}_{k}^{+})} + \sum_{k=1}^{K}\frac{-1}{d(\mathbf{x},\mathbf{x}_{k}^{-})} \Bigr).
\end{equation*}
In this work we used $K=3$. For finding nearest neighbors a Mahalanobis distance was used:
\begin{equation*}
d(\mathbf{x},\mathbf{y})=\sqrt{(\mathbf{x}-\mathbf{y})^{T}\mathbf{\Sigma}^{-1} (\mathbf{x}-\mathbf{y})},
\end{equation*}
where $\mathbf{\Sigma}^{-1}$ is inversion of covariance matrix calculated from training set. The choice of Mahalanobis distance is explained by the fact that components of feature vector $\mathbf{x}$ have non-zero correlation.

\subsection{Cross-validation process}
All classification experiments are performed using $k$-fold cross-validation (CV) method~\cite{Flach-2012}. Procedure starts with random mixing of the dataset, then it is divided into $k$ equal parts and during the following $k$ rounds one of the parts used as a test set and the remaining as training set. We used $k$=7. After all rounds (at each round new classifier is trained) the dataset  appears to be separated into two classes and accuracy of the ALS detector was assessed. CV process was repeated 40 times, then mean and standard deviation values for the performance metrics of classifier were calculated.

In order to measure the performance of the classifiers accuracy, sensitivity, specificity and averaged recall were calculated~\cite{Flach-2012}:
\begin{eqnarray}
\mathrm{Acc} &=&\displaystyle \frac{TP+TN}{TP+FP+FN+TN} \notag \\
\mathrm{Sens} &=& \displaystyle \frac{TP}{TP+FN} \notag\\
\mathrm{Spec} &=& \displaystyle \frac{TN}{TN+FP} \notag\\
\mathrm{R_{avg}}&=& \frac12 \bigl(\mathrm{Sens}+\mathrm{Spec}\bigr) \label{eq:R_avg} 
\end{eqnarray}
where $TP$, $TN$, $FP$, $FN$ -– the number of true positive, true negative, false positive and false negative results of detection, respectively. In this case, positive means a prediction that the voice sample is produced by a speaker with ALS. 

\section{Experiments}
\subsection{Dataset description}
The voice data used in this study was collected in Republican Research and Clinical Center of Neurology and Neurosurgery (Minsk, Belarus). A total of 54 speakers were recorded, with 39 healthy speakers (23 males, 16 females) and 15 ALS patients with signs of bulbar dysfunction (6 males, 9 females). The average age in the healthy group was 41.9 years (SD 16.3, Min 18, Max 82) and the average age in the ALS group was 57.7 years (SD 9.0, Min 40, Max 70). All the participants were asked to produce the sustained vowel /a/ at a comfortable pitch and loudness as constant and long as possible. The phonation was performed on one breath. The voice signals were recorded using smartphone with a headset (sample rate 44.1 kHz) and stored as 16 bit uncompressed PCM files. The records were manually edited to remove the beginning and ending of each utterance, removing the onset and offset effects in these parts of each utterance. Average duration of the samples is 4.1 s. 

Given that mean age for healthy group was about 17 years younger than ALS patients, we applied linear regression technique to remove age effect using the data of the healthy group. The correction was applied to the data of healthy speakers and ALS patients.

The voice database and Matlab tools used for the voice analysis are available in public GitHub repository\footnote{https://github.com/Mak-Sim/Troparion/tree/master/SPA2019}.

\subsection{Feature statistics}
\begin{figure}[t]
  \centering
  \includegraphics[width=\linewidth]{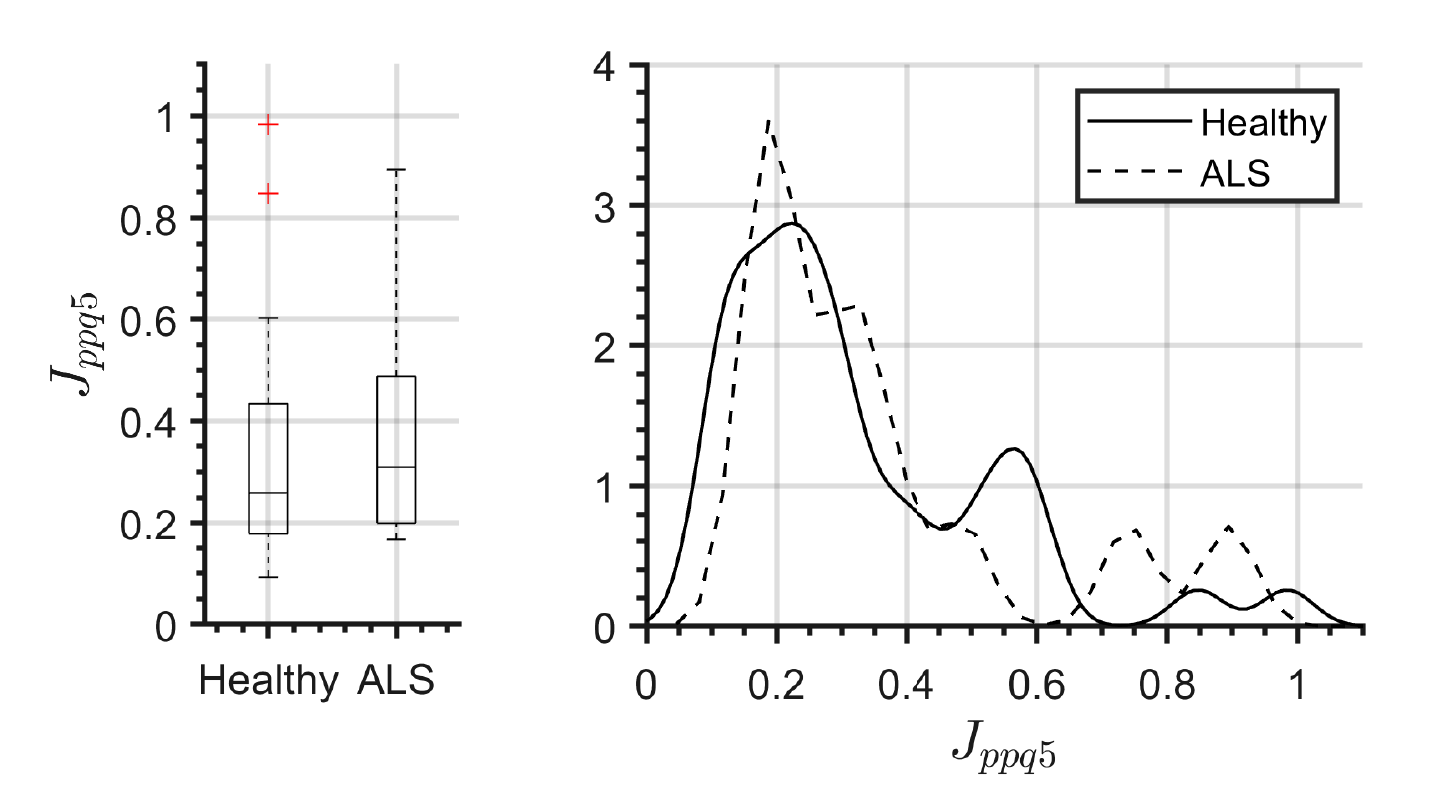}
  \caption{Box plot and probability densities of jitter measure}
  \label{fig:ppq5}
\end{figure}

In order to visualize the parameters of voice in HC and ALS groups we calculate statistics for several features. 
Figure~\ref{fig:ppq5} shows the box plot and distribution estimated using the Gaussian kernel density method for jitter measure. 
It can be seen that in ALS voices does not observed significant increasing of jitter compared to HC.

\begin{figure}[t]
  \centering
  \includegraphics[width=\linewidth]{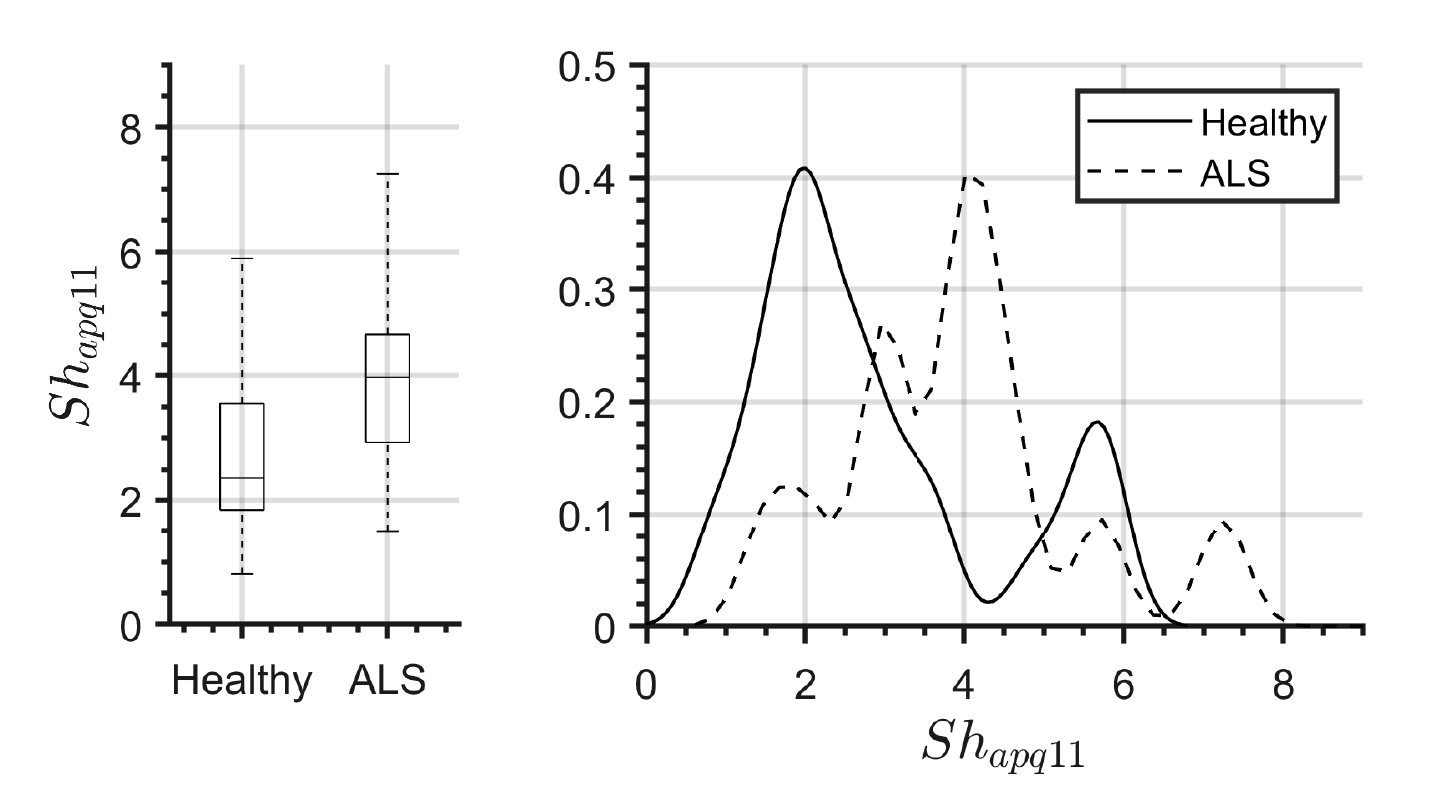}
  \caption{Box plot and probability densities of shimmer measure}
  \label{fig:apq11}
\end{figure}

\begin{figure}[t]
  \centering
  \includegraphics[width=\linewidth]{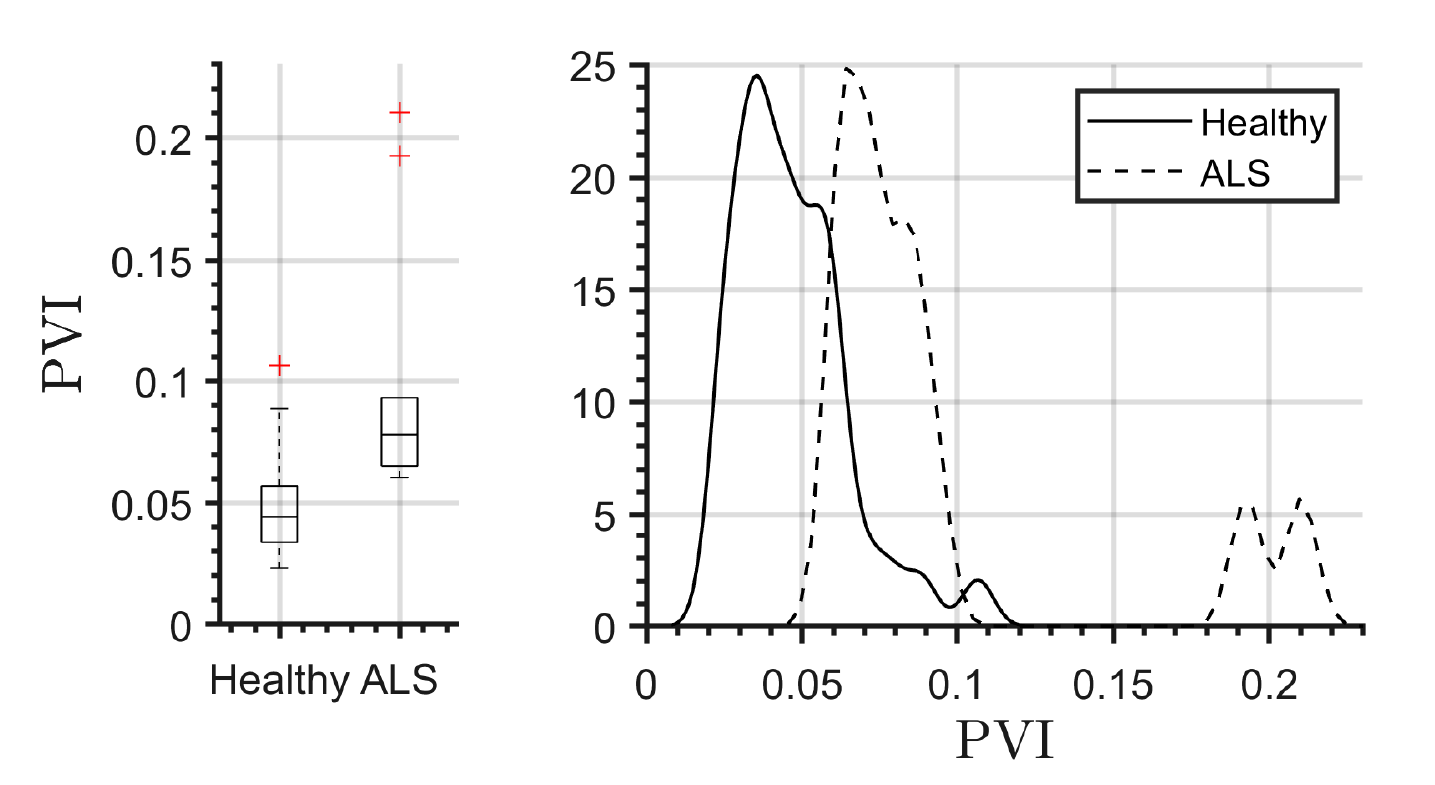}
  \caption{Box plot and probability densities of PVI}
  \label{fig:pvi}
\end{figure}

Fig.~\ref{fig:apq11} represents the box plot and kernel density estimation for shimmer measure that shows significant difference between healthy and ALS groups. 
In Fig.~\ref{fig:pvi} the same statistics for pathology vibrato index (PVI) is given. The $\mathrm{PVI}$ shows good separation capability.
A visual analysis of similar graphs for jitter parameters did not revealed a significant difference between the ALS patients voices compared to HC.


\subsection{Classification results and discussion}
We performed ALS classification for voice samples using following three feature sets: 1)~shimmers/jitters extracted using PRAAT~\cite{Boersma-02}; 2)~shimmers/jitters extracted using the proposed method; 3)~the same as in previous case but with PVI. 

Since the dimensionality $D$ of the  feature sets was low ($D$=7 for 1st and 2nd and $D$=8 for 3rd) we have performed exhaustive search through all possible feature subset to find the best one.

In the tables~\ref{tab:LDA_class} and~\ref{tab:kNN_class} for we used following notation: $J_1$ for $J_{loc}$, $J_3$ for $J_{rap}$, $J_5$ for $J_{ppq}$, $S_1$ for $S_{loc}$, $S_L$ for $S_{apqL}$.

Table~\ref{tab:LDA_class} shows the results of classification based on LDA\footnote{Accuracy, sensitivity and specificity are given in the form mean $\pm$ standard deviation and are out of sample computed using 7-fold cross validation with 40 iteration. Averaged recall is calculated using (\ref{eq:R_avg}) by taking mean sensitivity and specificity.}. For each feature set we show three best feature subsets. The results were ranked according to $\mathrm{R_{avg}}$ measure because our goal was to get classifier with both high specificity and sensitivity. It can be seen that best results are obtained for third feature set, that contained PVI. 

\begin{table}[th]
  \caption{LDA based classification results}
  \label{tab:LDA_class}
  \centering
  \begin{tabular} {c c c c c}
    \toprule
	\textbf{Features} & $\mathrm{R_{avg}}$ (\%) & $\mathrm{Acc}$ (\%) & $\mathrm{Sens}$ (\%) &  $\mathrm{Spec}$ (\%)	\\ 
    \midrule
    & \multicolumn{4}{c}{\it PRAAT features \vspace{4pt}}\\
$[S_5\;S_{11}]$ & 83.1 & 81.1$\pm$1.9 & 83.0$\pm$5.6 & 83.1$\pm$1.4\\
$[J_3\;J_5\;S_1\;S_5\;S_{11}]$ & 81.4 & 82.8$\pm$2.0 & 78.3$\pm$4.7 & 84.6$\pm$2.6\\
$[J_1\;J_5\;S_1\;S_5\;S_{11}]$ & 81.2 & 83.0$\pm$7.8 & 77.3$\pm$5.8 &  85.1$\pm$2.8\vspace{4pt}\\
	& \multicolumn{4}{c}{ \it Features based on WM with PC \vspace{4pt}}\\
$[J_1\;J_5\;S_3\;S_{11}]$ & 86.0 & 86.4$\pm$2.3 & 85.0$\pm$3.3 & 87.0$\pm$2.6\\	
$[J_5\;S_1\;S_5\;S_{11}]$ & 84.9 & 85.0$\pm$1.9 & 84.7$\pm$3.8 &  85.1$\pm$1.9\\
$[J_3\;J_5\;S_5\;S_{11}]$ & 84.4 & 86.5$\pm$1.6 & 79.8$\pm$1.8 &  89.0$\pm$2.0 \vspace{4pt}\\
	& \multicolumn{4}{c}{\it Features based on WM with PC + PVI \vspace{4pt}}\\
$[S_1\;S_3\;S_{11}\;\mathrm{PVI}]$ & {\bf 89.5} & {\bf 90.7$\pm$1.7} & {\bf 86.7$\pm$0.1} & {\bf 92.2$\pm$2.3}\\	
$[J_5\;S_1\;S_3\;S_{11}\;\mathrm{PVI}]$ & 88.0 & 89.0$\pm$2.2 & 85.8$\pm$2.7 &  90.2$\pm$3.0\\
$[J_1\;S_1\;S_3\;S_{11}\;\mathrm{PVI}]$ & 88.0 & 88.9$\pm$2.2 & 86.0$\pm$2.0 &  90.0$\pm$2.8\\	
    \bottomrule
  \end{tabular}  
\end{table}

The classifiers based on the PRAAT features have lower characteristics than classifiers that used jitters and shimmers extracted with proposed method.

Table~\ref{tab:kNN_class} shows the results of classification based on k-NN approach.

\begin{table}[th]
  \caption{kNN based classification results}
  \label{tab:kNN_class}
  \centering
  \begin{tabular} {c c c c c}
    \toprule
	\textbf{Features} & $\mathrm{R_{avg}}$ (\%) & $\mathrm{Acc}$ (\%) & $\mathrm{Sens}$ (\%) &  $\mathrm{Spec}$ (\%)	\\ 
    \midrule
    & \multicolumn{4}{c}{\it PRAAT features \vspace{4pt}}\\
$[J_3\;S_1\;S_3\;S_5]$ & 76.0 & 84.6$\pm$2.2 & 56.8$\pm$4.8 & 95.3$\pm$2.1\\
$[J_3\;S_1\;S_1\;S_3]$ & 75.8 & 84.2$\pm$1.7 & 56.8$\pm$4.3 & 94.7$\pm$2.0\\
$[J_1\;J_3\;S_1\;S_{11}]$ & 75.3 & 85.2$\pm$1.0 & 53.2$\pm$1.1 &  {\bf97.5$\pm$1.5}\vspace{4pt}\\
	& \multicolumn{4}{c}{ \it Features based on WM with PC \vspace{4pt}}\\
$[J_1\;J_3\;J_5\;S_5\;S_{11}]$ & 81.0 & 87.3$\pm$2.1 & 66.8$\pm$4.9 & 95.1$\pm$2.2\\	
$[J_3\;J_5\;S_5\;S_{11}]$ & 80.9 & 86.1$\pm$2.7 & 69.3$\pm$6.9 &  92.6$\pm$2.8\\
$[J_3\;J_5\;S_1\;S_3\;S_{11}]$ & 80.2 & 86.3$\pm$2.4 & 66.7$\pm$6.4 &  93.8$\pm$2.2 \vspace{4pt}\\
	& \multicolumn{4}{c}{\it Features based on WM with PC + PVI \vspace{4pt}}\\
$[J_1\;J_5\;\mathrm{PVI}]$ & {\bf 86.9} & {\bf 91.6$\pm$2.3} &  76.3$\pm$5.8 &  97.5$\pm$1.7\\	
$[J_3\;\mathrm{PVI}]$ & 86.5 & 90.5$\pm$1.5 & {\bf77.7$\pm$5.1} &  95.4$\pm$1.0\\
$[J_1\;\mathrm{PVI}]$ & 86.3 & 90.3$\pm$1.1 & 77.2$\pm$3.3 &  95.4$\pm$1.3\\	
    \bottomrule
  \end{tabular}  
\end{table}

A common feature of presented k-NN classifiers is that their specificity is much greater than sensitivity. Though the accuracy of the best k-NN classifier (91.6\%) is greater that best LDA classifier (90.7\%), the average recall of k-NN classifier (86.9\%) is considerably lower than in best LDA classifier (89.5\%). In general, the low sensitivity of the k-NN classifier  makes them less attractive for practical use.


\section{Conclusions}

This study presented improved voice analysis method for perturbation parameters estimation and new feature (PVI) for detecting bulbar dysfunction in ALS speakers.  PVI makes quantitative assessment of pathological changes of vibrato. Comparing data in tables~\ref{tab:LDA_class} and~\ref{tab:kNN_class} clearly indicates that adding PVI to voice perturbation parameters significantly improves the classification accuracy. The best achieved result is 90.7\% accuracy  (86.7\% sensitivity and 92.2\% specificity). 

The presented voice analysis methods may also be useful in discerning different levels of ALS severity. Future work will include other dysarthric-related disease to exclude a potential bias caused by common feature of dysarthric speech and focus on specific detection of ALS disease.

\bibliographystyle{IEEEtran}


\bibliographystyle{IEEEtran}
\bibliography{./IEEEabrv,mybib}

%


\end{document}